\documentclass[sigplan]{acmart} 
\renewcommand\footnotetextcopyrightpermission[1]{} 
\usepackage{mathrsfs}
\usepackage{graphicx}
\usepackage{minted}
\usepackage{xspace}
\usepackage{xcolor}
\usepackage{enumitem}
\usepackage{multirow}
\usepackage{booktabs}
\setlist{nosep}

\begin{document}

\title{Decoupled Attention Fusion: Accelerating RAG with Efficient KV Cache Reuse}
\author{Xiabao Wu\textsuperscript{1}, Wentao Liu\textsuperscript{2,1}, Yongchao Liu\textsuperscript{1}, Jiajun Zheng\textsuperscript{1}}
\affiliation{
  \institution{\textsuperscript{1}Ant Group, China; \textsuperscript{2}Southeast University, Nanjing, China}
  \city{}
\country{}}
\email{{wuxiabao.wxb,taoyang.lwt,yongchao.ly}@antgroup.com, zhengjiajun@gmail.com}

\maketitle
Retrieval-Augmented Generation (RAG) effectively mitigates hallucinations in Large Language Models (LLMs) but suffers from prohibitive Time-To-First-Token (TTFT) latency in long-context scenarios. Reusing pre-computed document KV caches addresses this but introduces a “distribution mismatch”, where offline caches lack the inter-document attention patterns required for coherent reasoning. CacheBlend~\cite{yao2025cacheblendfastlargelanguage} reduces recomputation via selective attention, but suffers severe accuracy degradation at longer contexts.
To address these challenges, we propose \textbf{Decoupled Attention Fusion} (DAF), a framework that maintains high accuracy while significantly reducing recomputation overhead. DAF decouples the attention process into three integrated stages: important-token self-attention to restore missing inter-document attention, question-document self-attention for standard inference, and a state fusion that concatenates their outputs to synthesize the final hidden states.
By decoupling these operations into dense patterns, DAF is natively compatible with Flash-Attention kernels, maximizing hardware utilization without requiring complex attention masks.
Experiments show that DAF delivers up to $2\times$ speedup over CacheBlend and $5.6\times$ over full recomputation with vLLM\cite{kwon2023efficient} on long-context benchmarks, without sacrificing accuracy.

\vspace{-3mm}
\section{Introduction}
Retrieval‑Augmented Generation (RAG) improves the factual reliability of large language models (LLMs) by grounding generation on retrieved external documents. However, when many documents are integrated into a single long context, inference becomes dominated by prefill latency—the Time‑to‑First‑Token (TTFT). This latency severely limits responsiveness and scalability in real‑time applications.

Unlike general long-context tasks, RAG features a static document corpus accessible beforehand. This allows for pre-computing KV caches offline to significantly reduce TTFT. This benefit comes at a cost: it suffers from “KV Cache Mismatch”,
as offline caches lack the essential inter-document attention during online joint inference. Such missing dependencies cause contextual representation shifts, leading to severe accuracy degradation.

Existing solutions attempt to mitigate this mismatch via different trade-offs. RAGCache~\cite{jin2024ragcacheefficientknowledgecaching} avoids accuracy loss by exhaustively caching document states at various sequence positions, but its brute-force approach incurs prohibitive storage overhead as the corpus scales. In contrast, the state-of-the-art CacheBlend~\cite{yao2025cacheblendfastlargelanguage} employs selective recomputation of "important" tokens to balance efficiency and precision. However, as the context length increases, its accumulated approximation errors lead to noticeable degradation in retrieval and generation quality, limiting its scalability in truly long-context RAG scenarios.

To address these challenges, we observe that the KV cache mismatch can be effectively rectified by performing self-attention solely among a subset of "important" tokens.
Based on this insight, we propose Decoupled Attention Fusion (DAF). DAF re-architects the prefill stage into three integrated components: (1) important-token self-attention to recover for inter-document attention; (2) question-document self-attention to maintain generative capability; and (3) a state fusion that concatenates the outputs from both to synthesize the final hidden states. By reformulating these decoupled operations into standard dense kernels, DAF maintains native compatibility with Flash-Attention. This enables peak hardware throughput and significantly lower latency while preserving high generation fidelity.

Extensive experiments show that DAF achieves up to $2\times$ speedup over CacheBlend, as well as $5.6\times$ speedup over full recomputation with vLLM, while sustaining comparable accuracy, demonstrating that principled attention decoupling can provide both efficiency and consistency in practical RAG systems.
\begin{figure}[t]
  \centering
  \includegraphics[width=1.0\linewidth]{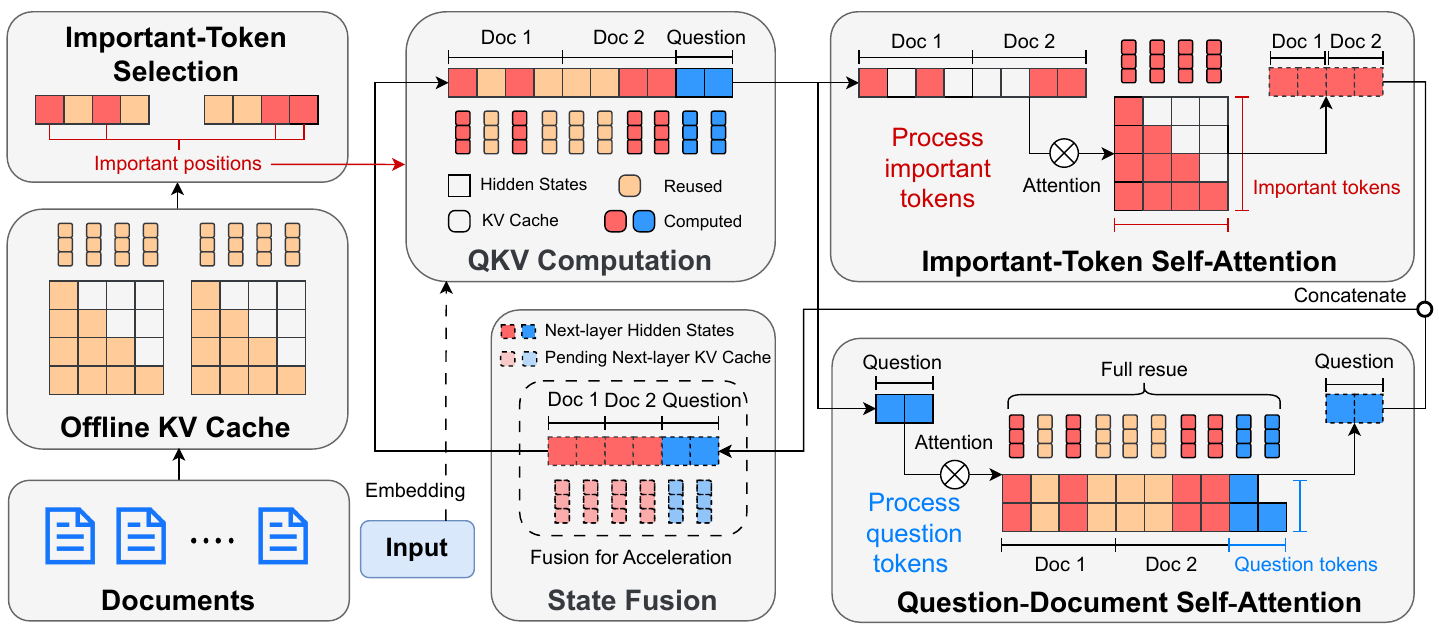}
  \caption{An overview of the proposed DAF framework.}
  \label{fig:overview}
\end{figure}
\vspace{-3mm}
\section{Methods: Decoupled Attention Fusion}
To address the mismatch between online and offline KV caches, we introduce DAF, a lightweight attention architecture that efficiently reconstructs missing dependencies, as shown in Fig.~\ref{fig:overview}.
Let $L_{d}$ denote the total number of tokens in all retrieved documents, $L_{q}$ the length of the question sequence ($L_{q}<<L_{d}$) in typical RAG settings, and $r$ ($r<1$) is the fraction of important tokens recomputed.
\begin{table}[!t] 
  \centering
  \small
  \caption{Speed comparison with vLLM on Qwen2.5-7B}
  \label{tab:qwen2.5-7b_perf}
  \setlength{\tabcolsep}{3pt}
  \resizebox{\columnwidth}{!}{
    \begin{tabular}{@{\extracolsep{\fill}} l c l c c c @{}}
      \toprule
      Dataset & Length & Method & Accuracy & TTFT (s) & Speedup \\
      \midrule
      \multirow{3}{*}{2WikiMultihopQA} & \multirow{3}{*}{6k} & vLLM & 0.389 & 0.450 & 1.0$\times$ \\
      & & CacheBlend & 0.363 & 0.250 & 1.8$\times$ \\
      & & DAF (Ours) & 0.393 & 0.250 & 1.8$\times$ \\
      \midrule
      \multirow{3}{*}{SAMSum} & \multirow{3}{*}{9k} & vLLM & 0.368 & 0.639 & 1.0$\times$ \\
      & & CacheBlend & 0.357 & 0.360 & 1.8$\times$ \\
      & & DAF (Ours) & 0.363 & \textbf{0.337} & \textbf{1.9$\times$} \\
      \midrule
      \multirow{3}{*}{RULER-qa1} & \multirow{3}{*}{30k} & vLLM & 0.744 & 2.32 & 1.0$\times$ \\
      & & CacheBlend & 0.645 & 1.10 & 2.1$\times$ \\
      & & DAF (Ours) & 0.742 & \textbf{0.82} & \textbf{2.8$\times$} \\
      \midrule
      \multirow{3}{*}{RULER-qa2} & \multirow{3}{*}{30k} & vLLM & 0.765 & 2.34 & 1.0$\times$ \\
      & & CacheBlend & 0.625 & 1.07 & 2.2$\times$ \\
      & & DAF (Ours) & 0.720 & \textbf{0.80} & \textbf{2.9$\times$} \\
      \midrule
      \multirow{3}{*}{\shortstack{LongBenchV2\\(Medium)}} & \multirow{3}{*}{136k} & vLLM & 0.288 & 23.30 & 1.0$\times$ \\
      & & CacheBlend & 0.180 & 8.370 & 2.8$\times$ \\
      & & DAF (Ours) & 0.250 & \textbf{4.160} & \textbf{5.6$\times$} \\
      \bottomrule
    \end{tabular}
  }
  \vspace{3mm}
\end{table}
\begin{enumerate}[leftmargin=*]
  \item \textbf{Important-Token Selection \& Self-Attention}
   To repair cross-document dependencies missing in independently encoded offline caches, we focus on a sparse subset of ``important tokens''.
    Unlike CacheBlend, which identifies these tokens based on KV deviation from a single layer, DAF aggregates signals across multiple layers (typically 2nd–4th) to mitigates layer-specific variance, ensuring robust selection. In self-attention phase, we gather these important tokens from all documents into a condensed sequence and compute self-attention mutually among them. This operation creates a global interaction field where pivotal information propagates across document boundaries, effectively reconnecting the isolated contexts while incurring only $\mathcal{O}(r^2 L_{d}^2)$ computational.

  \item \textbf{Question‑Document Self‑Attention}
    Positioned at the tail of the input sequence, the question tokens perform attention over the entire context. This operation generates an attention matrix of size $L_{q} \times (L_{d} + L_{q})$, where the Q vectors are derived from the question and the K/V vectors cover all preceding documents and the question. Thereinto, the document KV cache is no longer purely offline. Instead, the KV states for important tokens have already been updated during the QKV computation phase. By front-loading this selection logic, the current attention stage reuses the fully prepared hybrid KV cache. This design eliminates conditional checks (i.e., if-else branching), thereby avoiding the high overhead of thread divergence on GPUs and ensuring efficient parallel execution.

  \item \textbf{State Fusion}
    The outputs from attention paths are fused into the final hidden states to update the representation.
    This integration condenses sparse signals into a compact form, ensuring global contextual consistency while accelerating QKV computations in subsequent layers.
\end{enumerate}
\textit{\textbf{Complexity Analysis.}}
The efficiency improvement of DAF lies in reducing the effective coefficient of the quadratic computation term.
In CacheBlend, the recomputation involves all important tokens and question tokens over the full context, resulting in a complexity of approximately \mbox{$\mathcal{O}(rL_{d}^2 + L_{d}L_{q})$}. DAF further limits the interactions within a much smaller subset, where important tokens only attend to each other instead of the entire context, yielding a reduced cost of \mbox{$\mathcal{O}(r^2L_{d}^2 + L_{q}(L_{d}+L_{q}))$}. Although the dominant term still scales quadratically with $L_d$, the much smaller multiplier $r^2$ brings significant reductions in actual latency.

\textit{\textbf{Hardware Compatibility with Flash-Attention.}}
Unlike unified recomputation methods that often require complex, non-standard attention masks, DAF's decoupled components are structured as standard dense attention operations. This design allows DAF to natively leverage Flash-Attention kernels, ensuring high hardware utilization and translating theoretical complexity gains into significant wall-clock speedups.
\vspace{-0.5cm}
\section{Preliminary Results}

\textbf{Experimental Setup.} We evaluate DAF on Qwen2.5-7B using a single NVIDIA A100 (80GB) GPU under greedy decoding. We benchmark against vLLM (full-recomputation) and CacheBlend (state-of-the-art KV merging), fixing the important-token fraction at $r=0.3$ for fairness. Our evaluation spans diverse context scales: short-context multi-hop reasoning and summarization (2WikiMultiHopQA, SAMSum, 6k--9k tokens)~\cite{yao2025cacheblendfastlargelanguage}, and extreme long-context retrieval/QA (RULER~\cite{hsieh2024ruler}, LongBench-V2~\cite{bai2024longbench2}, up to 136k tokens). We measure prefill efficiency via TTFT and generation fidelity using standard task-specific metrics (e.g., Exact Match for QA, Rouge-L for summarization), which are harmonized as ``Accuracy'' in Table~\ref{tab:qwen2.5-7b_perf}.

\textbf{Performance and Accuracy}: From Table 1, DAF consistently outperforms CacheBlend in both speed and accuracy, with the advantage becoming more pronounced as the context length increases. Overall, DAF maintains significantly higher accuracy than CacheBlend in long-context tasks.

\textbf{Efficiency at Scale}: At 9k tokens, DAF and CacheBlend show similar performance, but at 130k tokens, DAF achieves a $2\times$ speedup over CacheBlend and is $5.6\times$ faster than vLLM. This confirms that DAF’s $\mathcal{O}(r^2L_d^2)$ complexity effectively suppresses the quadratic growth of recomputation costs.

\textbf{Limitations}: First, $r$ requires task-specific tuning for the accuracy-efficiency trade-off. Second, storing and managing massive KV caches remains an industrial-scale bottleneck even with high-speed SSDs, due to the prohibitive storage overhead of the corpus.


\bibliographystyle{unsrt}
\bibliography{ref}

\end{document}